\documentclass[letter,prx,twocolumn,10pt,aps]{revtex4-2}
\usepackage{graphicx}
\usepackage{amsmath}
\usepackage{natbib}
\usepackage{xcolor}
\usepackage{xr}
\makeatletter
\newcommand*{\addFileDependency}[1]{
\typeout{(#1)}
\@addtofilelist{#1}
\IfFileExists{#1}{}{\typeout{No file #1.}}
}\makeatother

\newcommand*{\myexternaldocument}[1]{%
\externaldocument{#1}%
\addFileDependency{#1.tex}%
\addFileDependency{#1.aux}%
}

\myexternaldocument{Supplemental_V2_current}

\begin{document}

\title{Jamming memory into acoustically trained dense suspensions under shear}
\author{Edward Y. X. Ong$^1$, Anna R. Barth$^2$, Navneet Singh$^2$, Meera Ramaswamy$^2$, Abhishek Shetty$^3$, Bulbul Chakraborty$^4$, James P. Sethna$^2$, Itai Cohen$^2$}
\affiliation{$^1$ Department of Applied Engineering and Physics, Cornell University, Ithaca, New York 14850, USA}
\affiliation{$^2$ Department of Physics, Cornell University, Ithaca, New York 14850, USA}
\affiliation{$^3$ Department of Rheology, Anton Paar, Ashland, Virginia 23005, USA}
\affiliation{$^4$ Department of Physics, Brandeis University, Waltham, Massachusetts 02453, USA}
\date{\today}

\begin{abstract}
Systems driven far from equilibrium often retain structural memories of their processing history. This memory has, in some cases, be shown to dramatically alter the material response. For example, work hardening in crystalline metals can alter the hardness, yield strength, and tensile strength to prevent catastrophic failure. Whether memory of processing history can be similarly exploited in flowing systems, where significantly larger changes in structure should be possible, remains poorly understood. Here, we demonstrate a promising route to embedding such useful memories. We build on work showing that exposing a sheared dense suspension to acoustic perturbations of different power allows for dramatically tuning the sheared suspension viscosity and underlying structure. We find that, for sufficiently dense suspensions, upon removing the acoustic perturbations, the suspension shear jams with shear stress contributions from the maximum compressive and maximum extensive axes that reflect or "remember" the acoustic training. Because the contributions from these two orthogonal axes to the total shear stress are antagonistic, it is possible to tune the resulting suspension response in surprising ways. For example, we show that differently trained sheared suspensions exhibit: 1) different susceptibility to the same acoustic perturbation; 2) orders of magnitude changes in their instantaneous viscosities upon shear reversal; and 3) even a shear stress that \textit{increases} in magnitude upon shear cessation. We work through these examples to explain the underlying mechanisms governing each behavior. Then, to illustrate the power of this approach for controlling suspension properties, we demonstrate that flowing states well below the shear jamming threshold can be shear jammed via acoustic training. Collectively, our work paves the way for using acoustically induced memory in dense suspensions to generate rapidly and widely tunable materials.       
\end{abstract}

\maketitle
\section{Introduction} 

Dense non-equilibrium systems have a memory of their history and can undergo large qualitative changes in their material property depending on the protocol under which they are prepared \cite{tant1981overview,jonason1998memory,pine2005chaos,corte2008random,mauro2009nonequilibrium,xiao2013modeling,keim2013multiple,fiocco2014encoding,prudnikov2016influence,keim2019memory,pashine2019directed,teich2021crystalline,lindeman2021multiple,arceri2021marginal,zhao2022ultrastable,zhao2022microscopic,vansaders2018strain,berthier2019rigidity}. For instance, the yield stress of a crystalline metal will increase under repeated deformation (work hardening) \cite{koehler1952nature,mott1952cxvii,kocks1976laws,sevillano1980large}. As another example, a granular material will densify when tapped and become harder to shear \cite{hong1994granular,lesaffre2000densification,slocombe2000densification,philippe2002compaction,richard2005slow,ribiere2007existence,iikawa2015structural}. In suspensions, numerous investigations have shown that shear induced memory can give rise to interesting behaviors including return point memory, coupling between hysterons, and increased yield strains \cite{keim2013multiple,keim2019memory,vaagberg2011glassiness,teich2021crystalline,galloway2022relationships,schwen2020embedding}. Memory in these studies is typically created through applying perturbations such as small amplitude oscillatory shear, which provides a controlled environment for studying the physics by limiting the extent to which the microstructure is disrupted. The downside of this controlled approach, however, is that the changes induced in the bulk material properties are often relatively modest in comparison to what should be possible. In particular, theoretical studies have shown that through reorganization of contact networks (tuning by pruning) it should be possible to generate up to 13 orders of magnitude change in the ratio of shear to bulk modulus \cite{goodrich2015principle,rocks2017designing,reid2018auxetic,hexner2018role,liarte2020multifunctional,rocks2017designing,reid2018auxetic,reid2019ideal,jin2022designing,pashine2023reprogrammable}. These studies suggest that if we can significantly alter the suspension contact network in a controlled fashion, we may be able to dial in or select dramatically altered suspension properties.   

Generating and retaining memory of large changes in contact networks in flowing systems, however, is challenging since the shear flow often reorganizes the systems and erases any memory of the training protocols. A promising but relatively unexplored system that could overcome this limitation is a shear jamming suspension, which is characterized by a rapid transition from a flowing state to a jammed state when driven beyond a critical applied shear stress  \cite{cates1998jamming,liu1998jamming,bi2011jamming,vaagberg2011glassiness,bertrand2016protocol,vinutha2016disentangling,luding2016so,han2016high,peters2016direct,romcke2021collision,henkes2016rigid}. Here, the suspension structure could be significantly modified while it is flowing and signatures of that structure could be remembered by rapidly shear jamming the suspension. In such systems, it may be possible to create a variety of shear jammed states with dramatically different material responses through different training protocols.

\begin{figure*}[ht!]
\centering
\includegraphics[width=1.95\columnwidth,trim=0.5cm 0.2cm 0cm 0cm, clip=false]{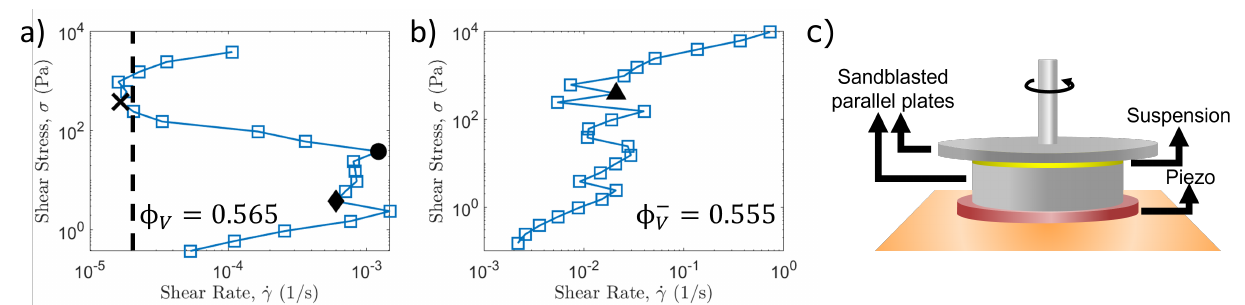}
\caption{\small{{Plots of shear stress versus shear rate for silica in dipropylene glycol at volume fractions of a) 0.565 and b) 0.555. For a), the ``X'' marker indicates a shear stress of 380Pa, which most of the experiments in the paper were performed at. The circle corresponds to a stress of 38Pa, applied for the protocol in Fig.~6b. The diamond corresponds to a stress of 3.8Pa, applied for the protocol in Fig.~S6 of the supplementary material. For b), the triangle corresponds to a stress of 380Pa, applied for the protocol in Fig.~6a. c) Schematic of the customized geometry used to apply both shear and acoustic perturbation.}}}
\label{fig:Setup}
\end{figure*}

We demonstrate that this approach is indeed feasible by applying acoustic perturbations to shear jamming suspensions. Previous studies have demonstrated that acoustic perturbations can significantly disrupt a suspension's microstructure to dethicken a shear thickened suspension \cite{sehgal2019using,sehgal2022viscosity}. Memory of the dethickening, however, is rapidly forgotten upon removal of the perturbation. By working with shear jamming suspensions, we can retain memory of this structural reorganization even after both shear and acoustic perturbations have been removed. Specifically, memory of the training induced by both shear and acoustic is stored in force networks oriented along both the maximum compressive and extensional axes. Signatures of this memory is then preserved when the suspension is jammed by the shear stress. We show that these antagonistic networks can be tuned to generate unexpected behaviors and large changes in the material properties of the suspensions.

\begin{figure*}[ht!]
\centering
\includegraphics[width=1.8\columnwidth,trim=0.3cm 0.1cm 0cm 0cm, clip=false]{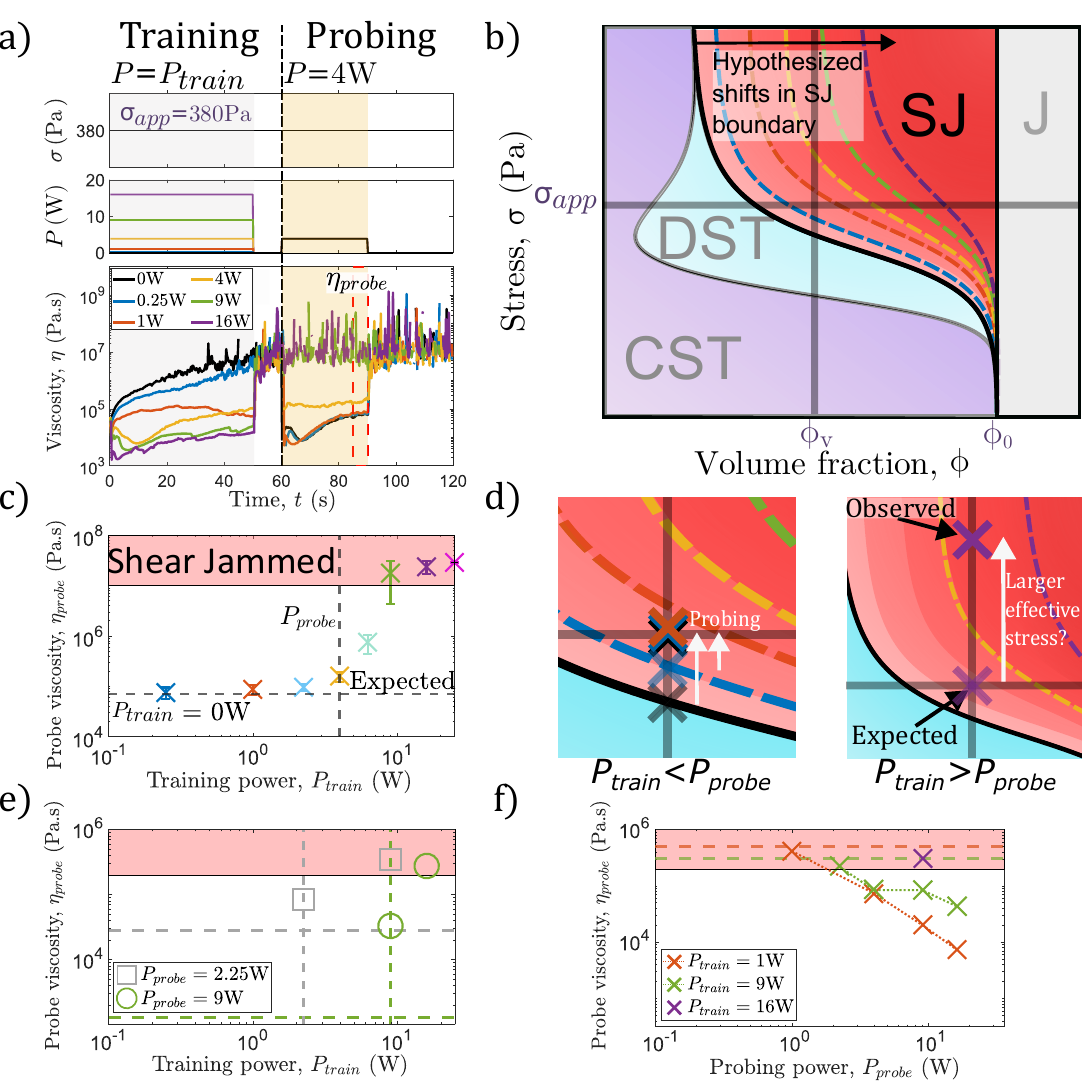}
\caption{\small{(a) Training (left) and probing (right) protocols plotted as shear stress (top), acoustic power (middle), and viscosity (bottom) versus time. During training, different acoustic powers  ($P_{train}$) are applied during the first 50~s (grey patch), as indicated by different colors. During probing, a constant acoustic power $P_{probe}$=4~W was applied for all $P_{train}$ from 60-90~s (yellow patch). A constant shear stress $\sigma=380$~Pa was applied for both training and probing. (b) Phase diagram of shear stress versus volume fraction ($\sigma-\phi$) based on our model framework. The fixed experimental stress, $\sigma_{app}=380$~Pa and volume fraction, $\phi_v=0.565$ are indicated by the vertical and horizontal grey lines. The isotropic jamming volume fraction is labelled as $\phi_0$. During training, larger $P_{train}$ shifts the shear jamming boundary farther away from the $P_{train}$=0~W shear jamming boundary (thick black solid line), as indicated by the hypothetical dashed curves which color follow the legend in (a). (c) Plot of the average probe viscosity $\eta_{probe}$ against $P_{train}$ during the probing period (red dashed region in (a). (d) Zoomed in views of the phase diagram. For $P_{train}<P_{probe}$ (left), the suspension either shear jams at the appropriately shifted shear jamming boundary (translucent black, blue crosses) or at $\sigma_{app}$ (brown cross) after training. During probing, the shear jamming line is shifted above $\sigma_{app}$, allowing states that were shear jammed below $\sigma_{app}$ to rearrange their microstructure and flow towards $\sigma_{app}$ (white arrows). For $P_{train}>P_{probe}$ (right), we expect $\sigma_{app}$ to limit how deeply the system can probe the shear jammed regime (translucent cross). Instead, we observe a much larger $\eta_{probe}$ and even jamming. indicating that the shear jammed state lies above the 4~W ($P_{probe}$) shear jamming boundary and well above $\sigma_{app}$ (opaque cross). Thus, acoustic training must embed some sort of structural memory which is not accounted for in a simple manner by the applied shear stress. We emphasize that the dashed SJ boundary and arrows in b) and d) are not measurements but illustrations based on our hypothesis. e),f) Plots of $\eta_{probe}$ against e) $P_{train}$ and against f) $P_{probe}$ for a cornstarch in water suspension with mass fraction $\phi_m=$0.57.}}
\label{fig:Testing}
\end{figure*}

\section{Experimental Methods}
We carry out the measurements on charge-stabilized silica (Angstrom Sphere) with a diameter of 2~\textmu m suspended in dipropylene glycol (Sigma Aldrich), with flow curves for volume fractions of $\phi_v=0.565$ and $\phi_v^-=0.555$ shown in Figs.~1a and 1b respectively. Unless otherwise stated, the volume fraction used is $\phi_v=0.565$ and the applied shear stress is $\sigma_{app}=380$~Pa (indicated by the ``X'' marker in Fig.~1a), considerably lower than the incipient stress for slip but still higher than the critical shear jamming stress of $\sim200$~Pa. The sedimentation timescale of the suspension is $\sim1000$~s, which is far longer than any experimental time window in the work presented here. Additional information on the suspension can be found in the supplementary material.

Measurements were performed using a modified Anton Paar MCR-702 stress-controlled rheometer with a piezoelectric disk bonded to the lower plate, as illustrated in Fig.~1c. The piezoelectric disk was purchased from APC International (material 841, 21mm diameter and 1.80mm thickness) and bonded via epoxy to an aluminum (6061-T6) bottom plate (19mm diameter and 8.57 mm thickness). The acoustic perturbations are generated by exciting the piezoelectric disk using a sine waveform with a frequency of 1.16MHz, over a peak to peak voltage range of 0V to 160V corresponding to an expected amplitude range of 0 to 24nm, which is much smaller than the gap size of 0.64mm. To minimize slip, we used sandblasted plates as the shearing surfaces. The rheometer has a measurement noise floor of $\sim 0.01$ Pa and a strain rate resolution on the order of $10^{-5}$ s$^{-1}$. The suspension is considered shear jammed when the shear rate oscillates around zero at a rate comparable to the instrument sampling frequency. The transition to the shear jammed state is visually striking on a logarithmic plot of viscosity versus time where smooth viscosity curves transition to abrupt large amplitude spikes as seen in Fig.~\ref{fig:Testing}a. 

Careful sample preparation, loading, and a consistent preshear protocol prior to and between measurements yielded highly reproducible results (see S.I. for more details on sample preparation and equipment characterization). The data shown in this paper was collected over 6 different samples with volume fraction $\phi_v=0.565$ for Figs.~2,3,4 and over 2 different samples with volume fractions 0.555 and 0.565 for Figs.~6a and 6b respectively. We note that the results shown here are highly reproducible and can be observed at a different volume fraction (Fig.~S7), using a different geometry (Fig.~S8) and even with a different suspension (Figs.~S9, S10, and S11), as discussed in the supplementary material.

\section{Measurement Protocols and Results}\label{Results}

\subsection{Training protocol} \label{train}
We systematically generate different shear jammed states via a training protocol as shown in Fig.~\ref{fig:Testing}a. The protocol involves shearing the suspensions at a constant shear stress, $\sigma = 380$~Pa (top panel), and applying acoustic perturbations with different amplitudes controlled by the acoustic power, $P_{train}$ (middle panel) for 50~s (gray band). We find that for low acoustic powers, $P_{train} = 0$, and 0.25~W (black and blue curves), the suspension viscosity (bottom panel) increases and eventually jams. As the training power is increased further, we find that the suspension remains fluid with a viscosity that stabilizes at a lower and finite value. Once the acoustic perturbations are switched off, all trained suspensions rapidly shear jam. 

Such behavior is reminiscent of a stress-volume fraction phase diagram commonly used to describe transitions between flow behaviors in sheared dense suspensions \cite{liu1998jamming,bi2011jamming,luding2016so,seto2019shear}, as shown in Fig.~\ref{fig:Testing}b. Here, CST refers to continuous shear thickening, DST to discontinuous shear thickening, SJ to shear jamming, and J to isotropic jamming. In the classical description, the suspension is expected to isotropically jam at an infinitesimally small shear stress beyond a volume fraction $\phi_0$ (grey shaded region). Below $\phi_0$, a sufficiently strong shear stress can induce an anistropic jammed state (red shaded region) for any volume fraction larger than $\phi_J$, as demarcated by the solid black phase boundary. For our protocol, a similar fluid-to-solid transition is observed, except here the shear stress and volume fraction are kept fixed and the training power now determines the final state of the suspension.

We focus on the shear jamming (SJ) boundary (black curve) between the flowing and rigid suspension states and its dependence on the training power, as shown in the schematic in Fig.~\ref{fig:Testing}b. Since the $P_{train} = 0$~W suspension shear jams, we deduce that the applied stress must be set above the shear jamming boundary. Since suspensions trained at higher $P_{train}$ flow, we deduce that the constant applied shear stress of $\sigma_{app}=380$~Pa is no longer sufficient to shear jam the suspension at the same volume fraction and thus the shear jamming line must have shifted above $\sigma_{app}$ i.e. closer to the boundary of the isotropic frictionless jamming phase (J), as indicated by the dashed lines. We note that similar shifts in the shear jamming phase boundary have been proposed for sheared dense suspensions under orthogonal oscillations \cite{ness2018shaken,ramaswamy2022incorporating}. Since the suspension for $P_{train} = 0.25$~W and 1~W are respectively jammed and flowing during the application of acoustic perturbations, we deduce that the applied stress $\sigma_{app}$ for our suspension volume fraction $\phi_v$ must lie between the shifted shear jamming boundaries for these powers (i.e. between the dashed blue and dashed brown curves). When the acoustic perturbation is switched off at t$=$50~s, the suspensions rapidly shear jam, which we interpret as a rapid recovery of the shear jamming boundary to its original state (black line). 

\subsection{Acoustic probing} \label{acoustic}
To probe how memory of the acoustic training alters the properties of the shear jammed states, we apply a constant acoustic perturbation with $P_{probe}$~=~4~W over t~$= 60-90$~s (yellow band in Fig.~\ref{fig:Testing}a). We find that under this acoustic probing, suspensions trained at $P_{train} < 4$~W fluidize with a viscosity that stabilizes at a finite value. In contrast, suspensions trained at much higher acoustic powers remain shear jammed.  We quantify this behavior by plotting the stabilized viscosity during probing, $\eta_{probe}$ (averaged over red dashed region in Fig.~\ref{fig:Testing}a) versus the training power, $P_{train}$ in Fig.~\ref{fig:Testing}c. Here, the horizontal dashed line indicates the viscosity of the $P_{train} = 0$~W suspension and the vertical line indicates the power of the acoustic perturbation used to probe all of the suspensions, $P_{probe} = 4$~W. These observations cannot be explained by simple shifting of the shear jamming boundary and indicate that acoustic training embeds structural memory in the sheared suspensions.

To understand why, we contextualize the data for the low and high $P_{train}$ regimes in in Fig.~\ref{fig:Testing}d based on the framework presented in Fig.~\ref{fig:Testing}b. For flowing states, the suspension microstructure is known to rearrange under shear to reach a non-equilibrium steady state (defined by time independent distributions of contact number and stress) determined by the applied shear stress, $\sigma_{app}$ \cite{peters2016direct,seto2019shear,edens2021shear}. When systems are driven into the jammed state, they form solid structures that balance the stress via nearly elastic deformations. This transition from a flowing state to a solid shear jammed state occurs when the applied stress is increased beyond the critical jamming stress, which is determined by the shear jamming boundary \cite{cates1998jamming,bi2011jamming,vinutha2016disentangling}. For $P_{train} < P_{probe}$ the measurements conform to these expectations.
Here, during training, the suspension either shear jams at the appropriately shifted shear jamming boundary or reaches the applied stress $\sigma_{app}$ (translucent black, blue, and solid brown crosses in Fig.~\ref{fig:Testing}d Left). 
Upon probing these systems at $P_{probe}$ =4~W, the shear jamming line shifts above $\sigma_{app}$, which fluidizes the systems, enabling the suspension to reach (white arrows) the non-equilibrium steady state determined by $\sigma_{app}$ (gray horizontal line). Similarly, we would expect that for suspensions trained at higher $P_{train}>P_{probe}$, the \textbf{constant} applied $\sigma_{app}$ would limit how deeply the systems is able to probe the shear jammed regime (translucent purple cross in Fig.~\ref{fig:Testing}d Right), and a similar fluidization should be observed when $P_{probe}$ =4~W is applied. However, in stark contrast, we observe a much larger $\eta_{probe}$ and even jamming. Given that $\sigma_{app}$ and the volume fraction both remained constant, this inconsistency indicates that some component of the suspension stress must lie above the 4~W shear jamming boundary and well above $\sigma_{app}$ (opaque purple cross on the right of Fig.~\ref{fig:Testing}d) to drive the suspension into a shear jammed state post training. The effect of acoustic training is thus to embed some sort of structural memory that results in a stress balance that enables such a stress component to exist under a constant $\sigma_{app}$.

We note that the key aspects of acoustic training and probing observed for the silica suspension can be reproduced using a suspension of cornstarch in water at a mass fraction of 0.57 (see Figs.~S9 and S10 in the Supplementary Information for more suspension and experimental information), suggesting that acoustic training can be used as a general method to tune the behavior of different types of suspensions. The response of the cornstarch suspension to two different probe powers, $P_{probe}=$2.25W and $P_{probe}=9$~W is shown in Fig.~\ref{fig:Testing}e. For both $P_{probe}$ values, we find that the average probe viscosity is larger for larger training power, with suspensions trained at sufficiently high $P_{train}$ remaining shear jammed after the probing power is switched on. These findings are consistent with what was observed for the silica suspension. In addition, we find that the threshold $P_{train}$ required for the suspension to remain shear jammed is larger for higher $P_{probe}$, and the probe viscosity for unjammed states is lower for higher $P_{probe}$. These observations were again reproduced when a sweep across $P_{probe}$ was conducted while keeping $P_{train}$ fixed, as shown in Fig.~\ref{fig:Testing}f.

Our results demonstrate that the shear jammed state remembers, or contains signatures of the acoustic training. In particular, the choice of training power determines whether the suspension will flow or shear jam under later acoustic probing at some $P_{probe}$. Since both the volume fraction and applied stress remained constant across the training and probing protocols, we hypothesize that the acoustic training changes the stress balance of the suspension state such that for the shear jammed states, a stress component lies above both $\sigma_{app}$ and the shifted shear jamming boundary at $P_{probe}$.


\begin{figure*}[ht!]
\centering
\includegraphics[width=2\columnwidth,trim=0.5cm 0.2cm 0.5cm 0.3cm, clip=false]{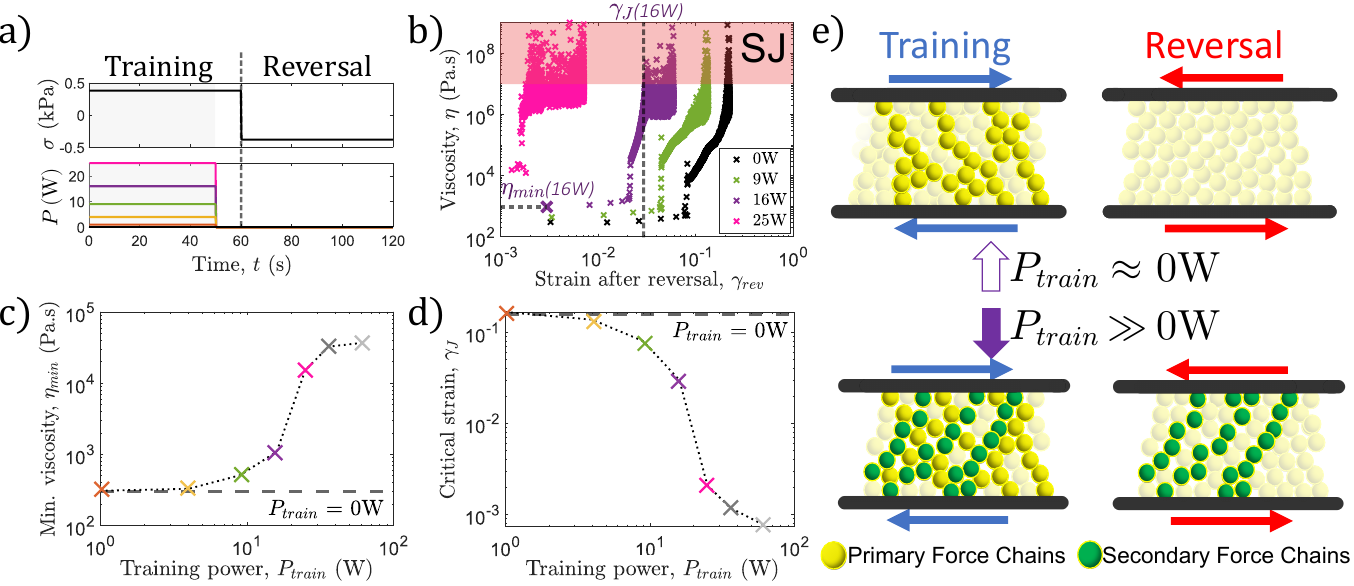}
\caption{\small{(a) Protocol for the shear reversal experiment plotted as stress (top) and acoustic power (bottom) against time. The first 60~s corresponds to training and the reversal occurs after 60~s. Colored plots correspond to different $P_{train}$. (b) Plot of the viscosity versus shear strain after reversal. Larger $P_{train}$ results in larger viscosities immediately after reversal ($\eta_{min}$) and shear jamming at lower critical strains ($\gamma_J$). As an example, $\eta_{min}$ and $\gamma_J$ are indicated by the dashed horizontal and dotted vertical line for $P_{train}=16$~W. (c) Plot of $\eta_{min}$ versus the training power. (d) Plot of $\gamma_J$ versus the training power. (e) Illustration of the deduced microstructures before and after reversal for low (top) and high (bottom) $P_{train}$. Yellow spheres indicate particles in primary force chains while green spheres indicate those in secondary force chains. For large $P_{train}$, significant secondary force chains exist prior to reversal which can resist the reversed shear flow and require little rearrangement to shear jam in the reversal direction.}}
\label{fig:Reversal}
\end{figure*}

\subsection{Shear reversal} \label{Reversal}
To gain further insights into our trained shear jammed states, the contact stress anisotropy of the suspension was characterized via a stress reversal protocol. Previous studies of continuously shear thickening and shear jamming suspensions undergoing shear reversal have demonstrated that the contact stresses responsible for shear thickening and jamming go to zero upon shear reversal as the particles which were initially compressed by the shear are now pulled apart \cite{gadala1980shear,lin2015hydrodynamic,ong2020stress,narumi2002transient,garat2022using}. As such, the viscosity after reversal, $\eta_{min}$, was observed to become significantly smaller (or finite) when compared to the shear thickened (or shear jammed) viscosity. With the contact stresses disappearing upon reversal, $\eta_{min}$ has been shown to be representative of the remaining hydrodynamic stresses and has been observed to remain nearly constant across different applied stresses before and after reversal.

The stress reversal protocol is shown in Fig.~\ref{fig:Reversal}a. Here, after training, we apply a shear stress with the same magnitude as the training stress, $\sigma_{app} = 380$~Pa, but in the reverse direction. The evolution of the suspension viscosity with strain after reversal is shown in Fig.~\ref{fig:Reversal}b, where only a subset of the curves were included for clarity. The figure shows that immediately after reversal, the suspension is unjammed and is characterized by a finite viscosity, $\eta_{min}$, that depends on $P_{train}$ (e.g. dashed grey line in Fig.~\ref{fig:Reversal}b indicating $\eta_{min} \sim 10^3$~Pa$.$s for $P_{train}=16$~W). Once a sufficiently large strain is built up, the suspension viscosity increases dramatically and eventually shear jams in the reverse direction at a critical strain, $\gamma_J$ (e.g. vertical grey dashed line for $P_{train}=16$~W). We find that both $\eta_{min}$ and $\gamma_J$ retain signatures of the acoustic training. 

To better visualize trends in these signatures, we plot $\eta_{min}$ and $\gamma_J$ versus $P_{train}$ in Fig.~\ref{fig:Reversal}c and Fig~\ref{fig:Reversal}d respectively. We find that $\eta_{min}$ increases by orders of magnitude with increasing training power, in stark contrast to the approximately constant $\eta_{min}$ found from typical shear reversal experiments \cite{gadala1980shear,lin2015hydrodynamic,ong2020stress,narumi2002transient,garat2022using}.
These measurements indicate that for higher training powers, there are many more secondary force networks built up in the jammed suspension that are aligned along the maximum extensional axis (i.e. perpendicular to the compressive axis normally responsible for shear jamming and DST). Upon reversal, these secondary networks align with the maximum compressive axis of the reversed shear direction and inhibit the flow Fig~\ref{fig:Reversal}c). Consistent with this picture, we find that after reversal much lower strains are necessary to jam suspensions trained at higher powers Fig~\ref{fig:Reversal}d). Our results suggest that larger $P_{train}$ generates shear jammed states that have significantly stronger secondary force chains along the extensional axis of shear prior to reversal.

These findings are summarized by the illustrations shown in Fig.~\ref{fig:Reversal}e. Here, the primary force networks responsible for shear thickening and jamming are depicted in yellow and are oriented along the shear compressive axis. Particles making up the secondary force networks along the maximal extension axis are depicted in green. At low training powers (top half), the suspension microstructure prior to reversal is strongly asymmetric, consisting mainly of primary force chains. Upon reversal, such force chains rapidly break up and the viscosity drops to the suspension viscosity prior to thickening. In addition, large strains are necessary to rearrange the particles in order to generate a shear jammed state. For high $P_{train}$ (bottom half), the suspension microstructure prior to reversal consists of both primary and secondary force networks (bottom left of figure). Upon reversal, even if the primary force networks break up, the secondary force networks can remain and resist the shear. This resistance results in large increases to the viscosity immediately after reversal. The presence of these secondary force networks also seeds or promotes an earlier transition to the shear jammed state after reversal. This picture of the suspension microstructure is strongly reminiscent of the ``fixed principle axes'' model, which attributes the ability of a shear jammed state to bear loads orthogonal to the compressive axis to a build-up of force chains along these orthogonal axes \cite{cates1998jamming,bouchaud1998models}.

In suspensions with a prominent secondary force network, the applied shear stress is balanced by the \textit{combined projection of the stresses} from networks oriented along both the compressive and extensional axes. Importantly, the primary and secondary force networks contribute to the shear stress in opposite directions -- the primary force networks resist the flow while the secondary force networks aid the flow. As such, the same applied shear stress can be accommodated by different combinations of shear stress projections from the competing force networks along the compressive and extensional axes. Therefore, to understand the behaviors of the differently trained jammed suspensions, the individual shear stress contributions from these competing force networks must be determined.

\begin{figure*}[ht!]
\centering
\includegraphics[width=1.6\columnwidth,trim=0cm 0.2cm 0cm 0.2cm, clip=false]{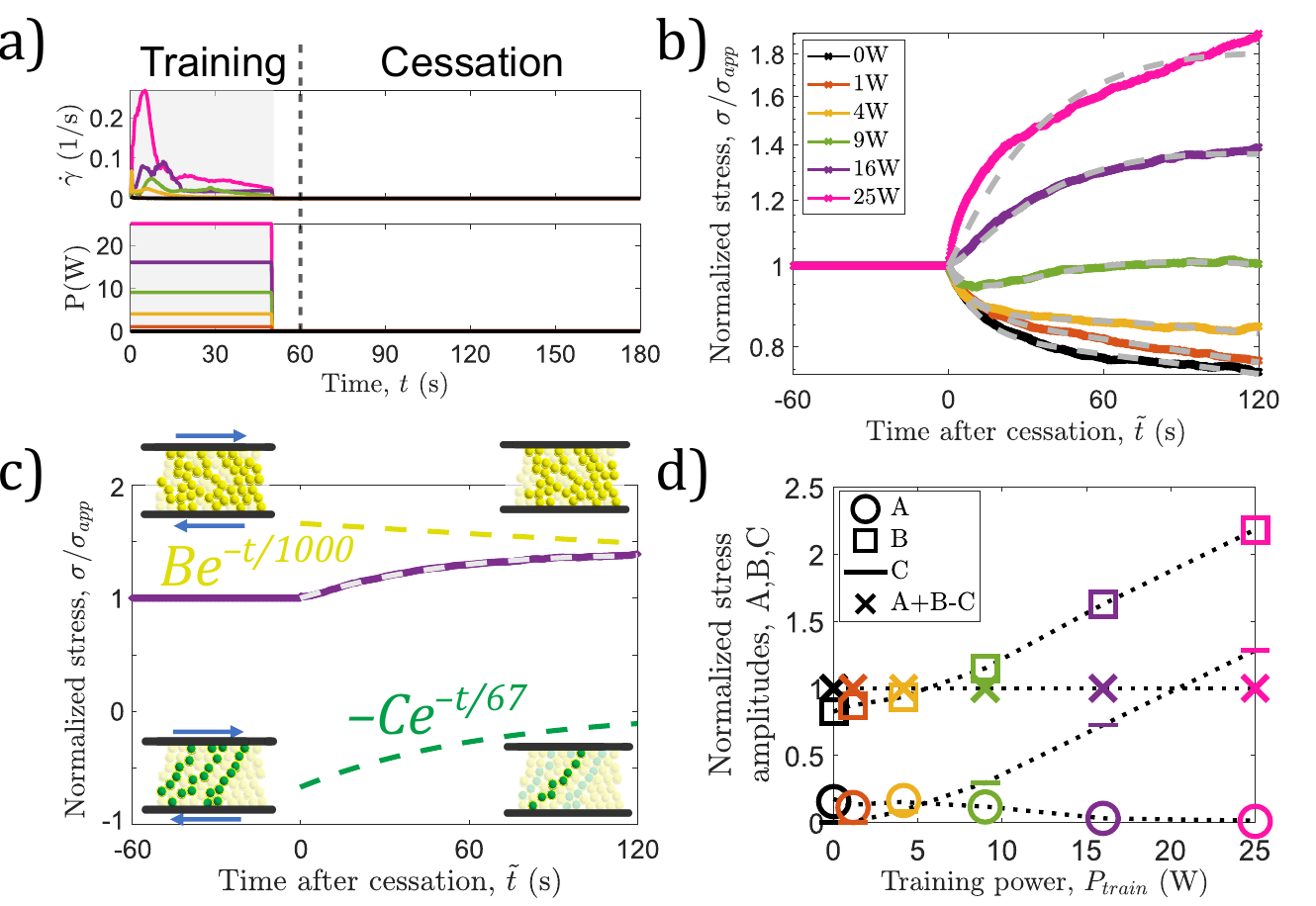}
\caption{\small{(a) Protocol for the rate cessation experiment where the shear rate (top) and acoustic power (bottom) are plotted versus time. The first 60s corresponds to training and cessation occurs from 60s onwards. Colored curves correspond to different $P_{train}$ as labeled in (b). (b) Plot of the instantaneous stress normalized by the applied shear stress versus time after cessation for different $P_{train}$. Dashed silver lines are fits to $Ae^{-\Tilde{t}/11} + Be^{-\Tilde{t}/1000} - Ce^{-\Tilde{t}/67}$ (c) A decomposition of the stress contribution from the $Be^{-\Tilde{t}/1000}$ and $-Ce^{-\Tilde{t}/67}$ terms used to fit the $P_{train}$=16W curve. The stress decays much more rapidly for the secondary force chains ($-Ce^{-\Tilde{t}/67}$ term, illustrated in green) compared to the primary force chains ($Be^{-\Tilde{t}/1000}$ term, yellow), resulting in the short time decay dynamics to be dominated by the relaxation of the secondary force chains. (d) Plot of the normalized stress amplitudes obtained as fit parameters for the fits in (b) versus $P_{train}$.}}
\label{fig:Cessation}
\end{figure*}

\subsection{Shear cessation} \label{Cessation}

We attempt to measure the shear stress contributions from the primary and secondary force networks by studying the suspension stress relaxation behavior via a rate cessation protocol as shown in Fig.~\ref{fig:Cessation}a. Here, the shear rate $\dot{\gamma}$ is fixed at zero after training, and the stress acting on the top plate while the suspension relaxes between its two stationary confining boundaries is measured. Previous studies of shear thickening and shear jamming suspensions undergoing shear cessation have found that the shear stress monotonically decreases after the shear rate has been set to zero \cite{maharjan2017giant,ianni2006relaxation,baumgarten2020modeling,barik2022origin}. In addition, multiple timescales are required to describe the purportedly exponential decay behavior for larger volume fractions, and the exact origins of these timescales remain an area of active research. For our study, we are agnostic of the exact physical origin of these timescales, but instead use shear cessation as a technique to back out the individual contributions from the primary and secondary stress networks, which is possible if the networks relax at sufficiently different rates.

Fig.~\ref{fig:Cessation}b shows a plot of the normalized instantaneous shear stress, $\sigma/\sigma_{app}$, versus the time after cessation, $\tilde{t}$, where $\sigma_{app}=$ 380Pa corresponds to the constant stress applied during training that is fixed across all cessation experiments. For P$_{train}\leq 1$~W, the stress decay is monotonic as expected. Strikingly, for larger $P_{train}$, the stress \textit{increases} during cessation. Motivated by our stress reversal results, we model the relaxation of the normalized stress as a competition between the decay of the force networks formed along the compressive and extensional axes. We find that we are able to reasonably fit all the data by modifying the typical sum of two exponential terms used in literature to include an additional negative exponential component that represents the relaxation of the antagonistic secondary force chain:

\begin{equation}
\sigma(\Tilde{t})/\sigma_{app} = Ae^{-\Tilde{t}/11} + Be^{-\Tilde{t}/1000} - Ce^{-\Tilde{t}/67}  
\label{eq:fit_func}
\end{equation}
with the constraint:
\begin{equation}
A+B-C=1
\label{eq:fit_constraints}
\end{equation}
and A,B,C $>$ 0. Here, the stress contribution from the primary force networks are captured by the first two terms \cite{ianni2006relaxation,maharjan2017giant,baumgarten2020modeling,barik2022origin}, and the stress contribution of the secondary force networks in the negative direction are captured by the third term. We note that there is no reason a priori to assume that the timescales for the decay should remain constant, and the decision to fix these timescales in our paper was mainly to minimize the total degree of freedom required to achieve a reasonably good fit to all the cessation curves. Our fit results reveal that for large $P_{train}$, where significant increases in the stress is observed after cessation, A~$\approx 0$ and the stress dynamics can be captured using only the B and C terms. To illustrate how the competition between force network relaxations can give rise to an increasing stress upon cessation, we plot the stress contributions from the $Be^{-\Tilde{t}/1000}$ and $Ce^{-\Tilde{t}/67}$ terms for $P_{train}=16$~W in Fig.~\ref{fig:Cessation}c. Since the stress decay dynamics are dominated by the faster relaxation of the secondary stress network (green and lower schematics), we observe a net gain in total stress (purple), even as the primary stress network is also decaying yellow and upper schematics). Consistent with this picture, we observe that at very long times the stress decays to zero.

\begin{figure*}[ht!]
\centering
\includegraphics[width=1.5\columnwidth,trim=0cm 0cm 0cm 0.1cm, clip=false]{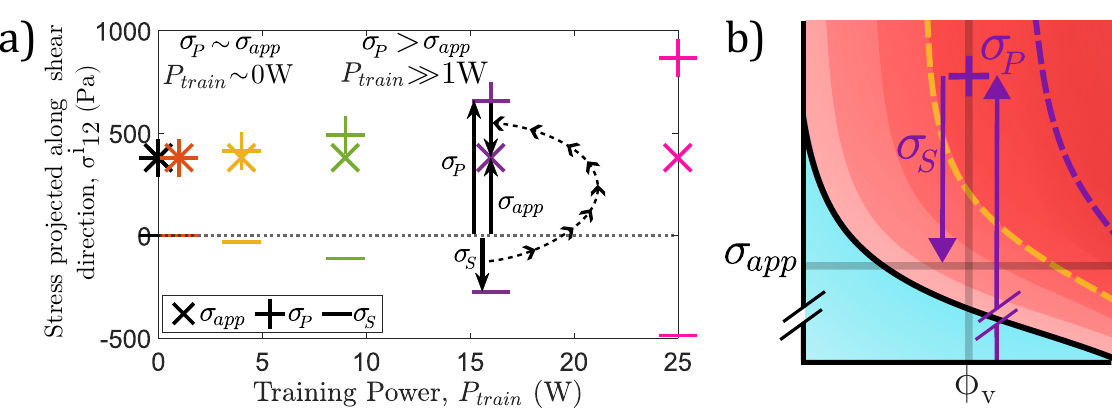}
\caption{\small{(a) Shear stress projections from the primary and secondary force chains calculated from Fig.\ref{fig:Cessation}d. The primary stress projection. $\sigma_{P}$, is represented by the ``$+$'' marker; the secondary stress projection, $\sigma_{S}$, by the ``$-$'' marker; and the (constant) applied stress, $\sigma_{app}$, by the ``X'' marker. The solid arrows are vectors indicating how the primary and secondary shear stress projections compete and sum to result in the applied shear stress for $P_{train}=16$~W. (b) The same vector representation is shown on the $\sigma-\phi$ phase diagram, illustrating how $\sigma_P$ (purple $+$ marker) can be above the 4W shear jamming boundary (yellow dashed line) even when the applied shear stress (horizontal grey line) is below the boundary at a given volume fraction (vertical grey line). The jammed structure along the primary direction and thus the suspension can remain shear jammed during probing under the 4W perturbation even though the applied shear stress remained unchanged. We note that the dashed SJ boundary and arrows in b) are not drawn to scale and are qualitative illustrations based on our hypothesis.}}
\label{fig:projection}
\end{figure*}

The values for the stress amplitudes $A$, $B$ and $C$ for all $P_{train}$ are potted in Fig.~\ref{fig:Cessation}d. We find that the coefficient $B$ that describes the main contribution of the primary force networks increases with training power and dominates the other terms. The coefficient $C$, is initially negligible but rapidly increases with $P_{train}$. This trend is consistent with the results in Fig.~\ref{fig:Reversal}c-e, which indicate an increased contribution from the secondary force networks in suspensions trained at larger $P_{train}$. Collectively, these cessation measurements and the accompanying analysis enable us to decompose the total stress into contributions from the primary and secondary force networks to explain the discrepancies described in Fig.~\ref{fig:Testing}d.  

\subsection{Shear stress projections}

Using the fits from the cessation measurements, we extract the stress contributions from primary and secondary force networks projected onto the shear axis, as shown in Fig.~\ref{fig:projection}a. The projection from the primary stress onto the shear direction, $\sigma_P= (A+B)\sigma_{app}$, is represented by the “+” marker; the projection from the secondary stress onto the shear direction, $\sigma_S=C\sigma_{app}$, by the “$-$” marker; and the (constant) applied stress, $\sigma_{app}=\sigma_P+\sigma_S$, by the “X” marker. We find that for low $P_{train}$, the stress contribution from the secondary network is negligible, $\sigma_S\sim0$, so that $\sigma_P \sim \sigma_{app}$. For large $P_{train}$, the contribution from the secondary network in the negative direction becomes significant. Consequently, the stress contribution from the primary force network \textit{must be greater than the applied stress}, $\sigma_P > \sigma_{app}$. The antagonistic relationship between the stress contributions $\sigma_P$ and $\sigma_S$, and how they sum up to $\sigma_{app}$ is represented vectorially (black arrows) for $P_{train}$=16W in Fig.~\ref{fig:projection}a. These results indicate that for systems with sufficiently developed secondary force networks, the contribution from the primary stress networks can be substantially larger than the applied stress. 

These large differences between $\sigma_P$ and $\sigma_{app}$ account for the behavior of suspensions trained at high $P_{train}$. We schematically illustrate the contributions of the primary and secondary force networks to the stress during the training and probing sequences in Fig.~\ref{fig:projection}b. During training the shear jamming boundary (solid black curve) shifts as illustrated by the purple dashed curve. Here, the training gives rise to a significant contribution from the secondary force network. To accommodate the applied stress, the contribution from the primary force network must be higher than $\sigma_{app}$ so that the sum of the contributions from the primary and secondary force networks match the applied stress (summation of purple arrows). Once the acoustic perturbations are removed, the suspension jams, preserving a memory of the force networks formed. During the probing period, if $\sigma_P$ is above the shear jamming boundary associated with $P_{test}=4$~W (dashed yellow line), the primary force networks remain shear jammed. 

We note that when compared to a shear jamming suspension at the same applied stress but at a lower volume fraction ($\phi_v=0.56$, see S.I. Fig.~S7), the acoustically induced increases in the magnitude of $\sigma_P$ and $\sigma_S$ are weaker for the same $P_{train}$. This difference in memory of the same training manifests as a larger threshold $P_{train}$ required to observe an increase in stress during shear cessation, alongside smaller changes in $\eta_{min}$ and $\gamma_J$ for a given $P_{train}$ for the less dense suspension. As such, we expect $\sigma_P$ and $\sigma_S$ to be potentially useful predictors of these behavioral changes induced by acoustic training.

\subsection{Inducing Solidification in Flowing States}

\begin{figure*}[ht!]
\centering
\includegraphics[width=1.9\columnwidth,trim=0cm 0cm 0cm 0.1cm, clip=false]{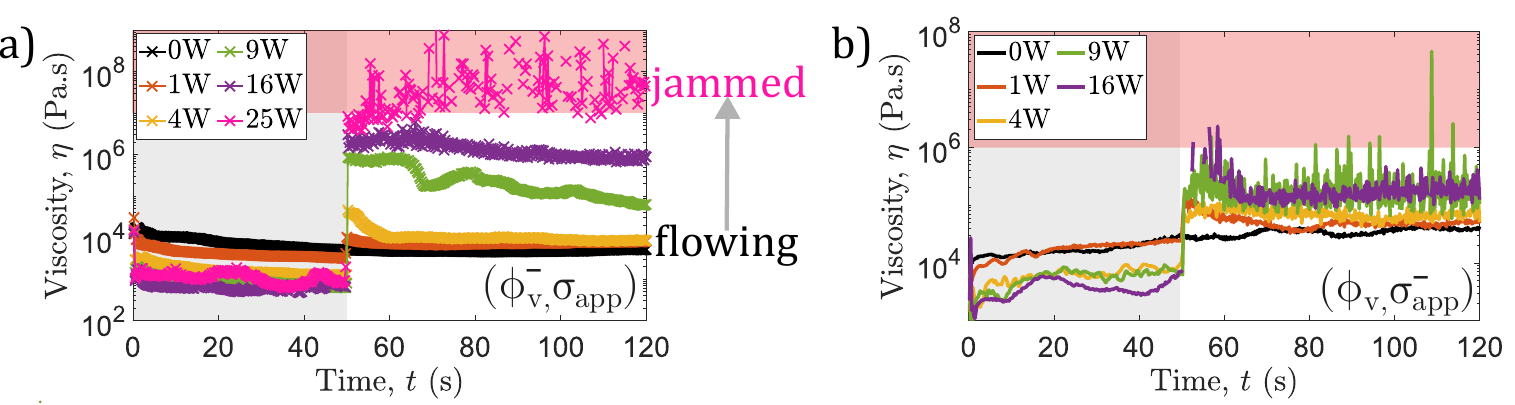}
\caption{\small{ Viscosity against time plot for suspensions with a) a volume fraction $\phi_v^-$ where $\phi_v^-=\phi_v- 1\%$, sheared at $\sigma_{app}$ (triangle marker in Fig.~1b) and b) a volume fraction of $\phi_v$ sheared at a stress $\sigma_{app}^-=\sigma_{app}/10$ (circle marker in Fig.~1a). The suspensions were trained for the first 60s (as in Fig.\ref{fig:Testing}a) and the applied shear stress of $\sigma_{app}$ for a) and $\sigma_{app}^-$ for b) was left on for an additional 60s. These flowing states shear jam when trained at sufficiently large $P_{train}$.}}
\label{fig:DST}
\end{figure*}

These results raise the intriguing possibility of training a flowing suspension that is well bellow the jamming line to produce a jammed state. We demonstrate that this is indeed possible by working with two suspensions: one at a 1\% lower volume fraction ($\phi_v^-$) sheared at $\sigma_{app}$ (triangle marker in Fig.~1b) and a second at the same volume fraction sheared at $\sigma_{app}/10$ ($\sigma_{app}^-$) (circle marker in Fig.~1a). Here, we train the two systems as described previously and probe their state by shearing with the same applied stress for an additional 60s after training. The viscosity versus time plots are shown in Fig.~\ref{fig:DST}. In both cases, when the suspensions are trained at $P_{train}=0$~W (black lines in Figs.~\ref{fig:DST}a,b), they are characterized by a finite steady viscosity. In other words, they flow. As $P_{train}$ is increased, we see the typical monotonic decrease in the viscosity when the acoustic perturbations are applied (gray shaded regions). When the perturbation is removed however, clear viscosity overshoots are observed. For weak $P_{train}$, these overshoots quickly decay back to the $P_{train}=0$~W viscosity. This decay timescale increases with $P_{train}$ until the system enters a shear jammed state as seen for $P_{train}=25$~W data in Fig.~\ref{fig:DST}a and for the $P_{train}=16$~W data in Fig.~\ref{fig:DST}b. We note that the order of magnitude lower stress applied in Fig.~\ref{fig:DST}b may have resulted in the relatively faster relaxation from the jammed state, as seen from the decrease in viscosity shortly after large fluctuations (signature of the suspension jamming) are observed. Nonetheless, we note that the transient behavior post training is consistent with Fig.~\ref{fig:DST}a prior to the viscosity relaxation. To further clarify the effect of stress on the acoustically induced memory in our suspensions, the same protocol was repeated for a suspension farther away from the shear jamming line, with the same volume fraction but at a stress 10 times smaller than in Fig.~\ref{fig:DST}b (i.e. $\phi_v=0.565$, $\sigma=3.8$~Pa, diamond marker in Fig.~1a), as shown in Fig.~S6 in the supplementary material. There, we find a rapid convergence of the viscosity to the $P_{train}=0$~W curve after the acoustic perturbation is removed for all $P_{train}$ applied. As such, acoustic training can clearly embed memory in flowing suspensions sufficiently near to the shear jamming line, even inducing solidification as demonstrated here. We find that for the ($\phi_v^-,\sigma_{app}$) state, the induced shear jamming can be preserved for a period at least $10$ times longer than the training interval (see Fig.~S5 in the supplementary material). While this effect does dissipate for systems driven sufficiently below the shear jamming line, the ability to shear jam flowing suspensions is nevertheless a remarkable demonstration of the impact that acoustic training can have on flowing suspensions.

\section{Discussion}
Our strategy for incorporating tunability in these networks involves two key steps: 1) the encoding of memory during the flowing state and 2) the rapid freezing of these network structures to preserve signatures of the encoding. Since these steps can be implemented in a variety of shear jamming materials using different memory encoding and preservation mechanisms, our strategy opens up a large material design landscape for exploration. For example, in shear thickening suspensions different shear protocols and perturbations, which can be multi-directional and/or time dependent, may be utilized to generate a variety of force networks during the fluid state that can be preserved via jamming \cite{lin2016tunable,leahy2017controlling,gaudel2017bulk,ness2018shaken,gibaud2020rheoacoustic,niu2020tunable,sehgal2022viscosity,martin2013driving,morillas2020magnetorheology,clemmer2021shear}. We expect that similar protocols would work for preserving memories in magneto-rheological and electro-rheological fluids. Additional mechanisms to preserve the suspension memory such as photopolymerization and rapid solidification can further extend our strategy to less dense suspensions and even to different systems such as foams and emulsion \cite{phillips1984photopolymerization,kaur2002photopolymerization,lavernia2010rapid}. Our general strategy for tuning the fluid state and leveraging a wide range of memory preservation mechanisms should enable us to generate a wide range of material properties. 

In shear thickening suspensions, such as system studied here, the build up of antagonistic force networks is a particularly potent mechanism for controlling the resulting material properties. Specifically, such antagonistic networks provide an additional degree of freedom for satisfying the applied force constraint and therefore greater control over the emergent material properties. Similar strategies are used in a variety of systems ranging from sophisticated temperature control for regulating flowering behavior in plants and homeostasis in human \cite{pose2013temperature,casal2019thermomorphogenesis,voronova20215} to muscular manipulation of limbs for rapid movement and stabilization \cite{jaric1995role,holt2003increased,wade2010joint,motter2018antagonistic,feng2022experimental}. 

Our results also demonstrate that the applied shear stress $\sigma_{app}$ is not sufficient to describe the state of the system. This observation is consistent with simulations and theoretical studies which find that a tensorial approach is required to fully characterize and predict suspension properties especially at high volume fractions, where the stress distributions become more isotropic \cite{sun2011constitutive,chacko2018shear,giusteri2021shear,seto2019shear,jin2021jamming}. The importance of antagonistic force chains also suggest a way to extend the universal scaling framework \cite{ramaswamy2021universal,ramaswamy2022incorporating} to memory-forming suspensions.  The lesson learnt is that frictional contacts along the compressive and dilational directions play very different roles, and therefore impact the shear-jamming process differently. The results presented here show how decomposition of the shear stress contributions into antagonistic components may be required to fully capture the behavior of these very dense suspensions even under simple shear. Such stress decomposition is likely to also be necessary to fully capture and understand more complicated flow behaviors induced by combinations of shear and compression or other biaxial and even triaxial flow manipulations. 

We note that our results may also draw a deep connection to studies on directed ageing and memory encoding in glasses, gels and granular materials \cite{pashine2019directed,hexner2020effect,keim2013multiple,keim2019memory,josserand2000memory}. In particular, the memory encoding and preservation processes presented here may be analogous to annealing, quenching and/or densification. Here, application of larger acoustic power during training drives the system into a more stable configuration with a larger part of the network characterized by a longer relaxation timescale. This training process is reminiscent of annealing, which allows particles to explore their energy landscape and settle in a more stable energy minima. Our experimental non-thermal acoustic perturbation protocol, like numerical Wolff cluster moves and swap Monte Carlo~\cite{grigera2001fast,fernandez2007optimized,ninarello2017models,wolff1989collective}, bypasses the usual relaxation barriers and allows exploration of otherwise challenging or inaccessible materials morphologies. One notable difference between our approach and these previous training protocols is that the training in our system is conducted while the suspension is sheared. This breaking of symmetry enables training of tensorial components of the stress, which are not as easily accessed in these previous protocols~\cite{pashine2019directed,hexner2020effect,josserand2000memory,keim2019memory,schwen2020embedding,teich2021crystalline,arceri2021marginal,galloway2022relationships}. 

While our current study is limited by a lack of imaging data, we emphasize that this very difficulty in obtaining useful imaging data for very dense suspensions further highlights the importance of the characterization protocols used in this paper. There is a growing number of studies that demonstrate that shear thickening and shear jamming in non-deformable particles are caused by a redistribution of frictional interactions between particles without significant rearrangement of the geometric topology \cite{lin2015hydrodynamic,ong2020stress,edens2021shear,kim2023stress}. As a result, it remains extremely challenging, if not impossible, for the spatial changes associated with these changes in frictional interactions to be resolved optically \cite{edens2021shear,pradeep2021jamming,papadopoulos2018network}, even with advanced image reconstruction algorithms that provides nanometer-scale resolution \cite{bierbaum2017light,leahy2018quantitative}. Given that direct optical measurements are unlikely to yield results, we have turned to alternative measurement techniques, such as the protocols used in this study, to characterize the suspension and through these characterization, deduce changes in the microstructure. Such deductions are possible because the structural changes are directly associated with frictional forces being redistributed \cite{edens2021shear,gameiro2020interaction}, and so the fact that we are able to dramatically change the response of the jammed suspension indicates that there is something different about the organization of the frictional interaction, which indicates structural changes. In addition, as new materials and tuning mechanisms continue to be introduced into the field of dense suspension, alternative non-imaging based methods to obtain additional information on the suspension which do not impose additional constraints on the setup or suspension (e.g. on refractive index matching, transparency of surface etc) such as the protocols shown here will become increasingly important. As such, we believe the striking results obtained from these protocols are highly valuable even without, or perhaps exactly because of the lack of microscopic data available for suspensions in this very dense regime. We note that a recently demonstrated experimental protocol \cite{blanc2023rheology} that involves carrying out shear reversal protocols at different angles to the direction of shear may enable us to resolve the contribution of the acoustic training to the full stress tensor, and will be pursued in a future study.

\section{Conclusion}
Our strategy to \textit{jam} memory of antagonistic force networks into suspensions shows how flow protocols can be used to \textit{engineer} suspensions with dramatically different material properties. While we have focused on acoustic training and shear jamming in this study, our strategy is broadly applicable to different flow tuning and structure preservation mechanisms, which we expect to be fertile ground for uncovering and engineering unique material properties. In combination with more sophisticated training, flow, and solidification protocols, it should finally be possible to generate structured fluids that can begin to take advantage of the predicted orders of magnitude changes in materials properties that can emerge from altering network connectivity and force transmission.

\section{Acknowledgement}
We thank Anton Paar for use of the Twin Drive MCR 702 rheometer through their VIP academic  research program and the Cohen group for their insightful suggestions. We also thank Cornell Center for Materials Research, an NSF MRSEC supported by DMR-1719875, for providing access to their facilities. This work is also supported by NSF CBET award numbers 1804963, 1509308 and 2228681 as well as DMR-1507607 and DMR-2327094. Anna Barth is supported by NSF DGE-2139899. Edward Ong is supported by the Agency of Science Technology and Research, Singapore NSS(PhD) award.

\bibliographystyle{ieeetr}

\end{document}